\documentclass[proceedings]{JHEP3}
\usepackage{amsfonts}
\usepackage{amsmath}
\usepackage{epsfig,multicol}

\newbox\mybox

\newcommand\fverb{\setbox\mybox=\hbox\bgroup\verb}
\newcommand\fverbdo{\egroup\medskip\noindent\fbox{\unhbox\mybox}\ }
\newcommand\fverbit{\egroup\item[\fbox{\unhbox\mybox}]}
\conference{Integrable models from $\mathcal{PT}$-symmetric deformations}
\abstract{We address the question of whether integrable models allow 
for $\mathcal{PT}$-symmetric deformations which preserve their intgrability. For this purpose we carry out the Painlev\'{e} test for $\mathcal{PT}$-symmetric deformations of Burgers and the Korteweg-De Vries equation. We find that the former equation allows for infinitely many deformations which pass the Painlev\'{e} test. For a specific deformation we prove the convergence of the 
Painlev\'{e} expansion and thus establish the Painlev\'{e} property
for these models, which are therefore thought to be integrable. The Korteweg-De Vries equation does not allow for deformations which pass the Painlev\'{e} test in complete generality, but we are able to construct a defective Painlev\'{e} expansion.}

\title{Integrable models from $\mathcal{PT}$-symmetric deformations}
\author{Paulo E.G. Assis and Andreas Fring \\
Centre for Mathematical Science, City University London, \\
Northampton Square, London EC1V 0HB, UK\\
E-mail: Paulo.Goncalves-De-Assis.1@city.ac.uk, A.Fring@city.ac.uk}

\input{tcilatex}

\begin{document}

\section{Introduction}

Classical as well as quantum mechanical models, which are invariant under a
simultaneous parity transformation $\mathcal{P}:x\rightarrow -x$ and time
reversal $\mathcal{T}:t\rightarrow -t$, can be deformed in a controlled
manner to produce new $\mathcal{PT}$-symmetric theories \cite%
{Bender:1998ke,Benderrev,Bender:2006tz,Ali23,Bender:2007gm,Josh,BBCF,AFKdV}.
The crucial feature of these models is that the $\mathcal{PT}$-symmetry can
be utilized to guarantee the reality of the energy spectra, which is due to
the fact that its operator realization is a specific example of an
anti-linear operator \cite{EW}. In contrast to standard textbook wisdom,
this means when the systems are Hamiltonian, they are non-dissipative
despite being non-Hermitian. An important question to answer in this context
is whether it is possible to deform models in a symmetry preserving manner.
Regarding supersymmetry, it was recently shown \cite{BBAF} that this is
indeed possible. Here we will focus on the question of whether this is also
accomplishable with regard to the symmetry underlying integrability. In
other words, do integrable $\mathcal{PT}$-symmetric models allow for
deformations which do not destroy the integrability? A positive answer to
this question will naturally lead to new integrable models. For some cases
partial results already exist \cite%
{Basu-Mallick:2000af,Basu-Mallick:2001ce,Milos,AF,AFActa,AFZ,BBCF,AFKdV,BBAF}%
. Here we will focus on two prototype models of integrable systems, the
Burgers equation and the Korteweg-deVries (KdV) equation. We will carry out
the Painlev\'{e} test for $\mathcal{PT}$-symmetric deformations of these
models, establish thereafter in some cases the Painlev\'{e} property and
draw conclusions about their integrability.

As there exist various notions and definitions about integrability, the
Painlev\'{e} test, the Painlev\'{e} property, etc, let us briefly indicate
which ones we are going to adopt in this manuscript. To start with, there is
clearly no doubt that integrability is an extremely desirable property to
have in a physical system, as it usually leads to exact solvability rather
than to mere perturbative results. In the context of 1+1 dimensional quantum
field theories the notion of integrability is usually used synonymously to
the factorization of the scattering matrix, where the latter can be achieved
simply by making use of one non-trivial charge \cite{Witten}. Unlike as in
most scenarios when one compares quantum and classical theories, the latter
appear to be more complicated in this particular regard. In classical
systems the definitions of integrability are much more varied and
non-uniform. A common notion is so-called Liouville integrability, which
assumes for a system with $N$ degrees of freedom the existence of $N$
analytic single valued global integrals of motion in involution. The
equations of motion are then separable and exact solutions can be obtained,
at least in principle. Focussing on differential equations, as we do in this
paper, one calls them integrable when, given a sufficient amount of initial
data, they are solvable via an associated linear problem. The problem with
all these definitions is that one does not know a priori whether a system is
integrable or not without having computed all integrals of motion, mapped
the problem to a linear one or actually solved the equations of motion. A
general method to identify integrable models before this, often very
difficult, task is completed does not exist. The closest one may get to such
a method is to check whether the system possesses the Painlev\'{e} property.
One can then assume that the Painlev\'{e} property implies integrability in
the above specified sense, albeit this connection is not rigorously proven.
To make matters worse, there exist even definitions which include the notion
of integrability into the definition of the Painlev\'{e} property \cite%
{Conte}.

The concept of the Painlev\'{e} property can be traced back more than a
century to the original investigations of Painlev\'{e} et al. \cite{Painor},
who set out to construct new functions from the solutions of ordinary
differential equations (ODE). The notion of a function implies immediately
that the solutions one is seeking ought to be single valued, which leads to
a natural definition: An ODE whose (general) solutions have no movable%
\footnote{%
Movable means that the solution depends on the initial values.} critical%
\footnote{%
A critical singularity is multivalued in its neighbourhood.} singularities
is said to possess the (generalized) Painlev\'{e} property \cite{Conte,
Gramma,Kruskal}. The classification of possible solutions to this problem
can be organised into equivalence classes obtained from linear fractional (M%
\"{o}bius) transformations and has been completed only to some degree. It is
proven that all linear ODE posses the Painlev\'{e} property, first order
algebraic nonlinear equations lead to Weierstrass functions and second order
algebraic nonlinear equations lead to the famous six Painlev\'{e}
transcendental functions. The classification of algebraic ODEs with Painlev%
\'{e} property of order greater than two is still an open problem, albeit
some partial results exist \cite{Chazy,Bureau,Cos}.

The situation is somewhat less structured for partial differential equations
(PDE). Extrapolating the previous notions one defines: A PDE whose solutions
have no movable critical singularities near any noncharacteric\footnote{%
On a characteristic manifold we can not apply Cauchy's existence theorem and
therefore we do not have a unique solution for a given initial condition.}
manifold is said to possess the Painlev\'{e} property. In general this is
difficult to establish, however, there exists a more applicable necessary,
albeit not sufficient, condition for a PDE to possess the Painlev\'{e}
property, which was developed by Weiss, Tabor and Carneval \cite{Pain1} and
is usually referred to as the Painlev\'{e} test. This method is extremely
practical \ and can be carried out in a very systematic fashion. Roughly
speaking the main idea is that one expands the solution for a PDE (or ODE)
in a power series starting with some single valued leading order terms. In
case the series can be computed and involves as many free parameters as the
order of the PDE then it is said that the PDE passes the Painlev\'{e} test.
In order to extrapolate from the Painlev\'{e} test to the Painlev\'{e}
property one should also establish the convergence of the series, which,
however, has been carried out only in very rare cases.

For our purposes the relation between the Painlev\'{e} property (test) and
integrability is the most interesting. Ablowitz, Ramani and Segur \cite{ARS}
conjectured almost thirty years ago: Any ODE which arises as a reduction of
an integrable PDE, possibly accompanied by a variable transformation,
possesses the Painlev\'{e} property. To this day this conjecture has not
been proven rigorously, but is supported by a huge amount of evidence. On
one hand one has verified this property for almost all known integrable PDEs 
\cite{Pain1,Pain2,Hlava,Josh,KC} and in turn, which is more impressive, one
has also used it to identify new integrable ODEs \cite{BSV,DGR}. The latter
is what we hope to achieve in this manuscript.

In summary, we will adopt here the logic that a PDE which passes the Painlev%
\'{e} test and whose Painlev\'{e} expansion converges also possesses the
Painlev\'{e} property. We take this as a very good indication that the
system is integrable.

We briefly explain the deformation procedure in section 2 and carry out the
analysis for Burgers and the KdV equation in subsection 2.1 and 2.2,
respectively. We state our conclusions in section 3.

\section{$\mathcal{PT}$-symmetrically deformed integrable models}

Given a $\mathcal{PT}$-symmetric PDE as a \ starting point, we adopt the
deformation principle of \cite{BBCF,AFKdV,BBAF} to define new $\mathcal{PT}$%
-symmetric extensions of this model by replacing ordinary derivatives by
their deformed counterparts 
\begin{equation}
\partial _{x}f(x)\rightarrow -i(if_{x})^{\varepsilon }=:f_{x;\varepsilon }%
\text{\qquad \qquad with \ }\varepsilon \in \mathbb{R}.  \label{a}
\end{equation}
Clearly the original $\mathcal{PT}$-symmetry is preserved. In general the
deformations will continue real derivatives into the complex plane, unless $%
\varepsilon =2n-1$ with $n\in \mathbb{Z}$. We do not make use here of the
possibility to deform also the higher derivatives via the deformation (\ref%
{a}), i.e. replacing for instance $\partial _{x}^{2}f(x)$ by $%
f_{x;\varepsilon }\circ f_{x;\varepsilon }$, but simply define them as
successive action of ordinary derivatives on one deformation only 
\begin{equation}
\partial _{x}^{n}f(x)\rightarrow i^{\varepsilon -1}\partial
_{x}^{n-1}(f_{x})^{\varepsilon }=\partial _{x}^{n-1}f_{x;\varepsilon
}=:f_{nx;\varepsilon }.
\end{equation}
This deformation preserves the order of the PDE. We can now employ this
prescription to introduce new $\mathcal{PT}$-symmetric models.

\subsection{Painlev\'{e} test for the $\mathcal{PT}$-symmetrically deformed
Burgers' equation}

Burgers' equation is extensively studied in fluid dynamics and integrable
systems, as it constitutes the simplest PDE involving a nonlinear as well as
a dispersion term 
\begin{equation}
u_{t}+uu_{x}=\sigma u_{xx}.  \label{Burgers}
\end{equation}
Obviously equation (\ref{Burgers}) remains invariant under the
transformation $t\rightarrow -t,x\rightarrow -x$, $u\rightarrow u$ and $%
\sigma \rightarrow -\sigma $. Taking the constant $\sigma $ to be purely
imaginary, i.e. $\sigma \in i\mathbb{R}$, this invariance can be interpreted
as a $\mathcal{PT}$-symmetry, which was also noted recently by Yan \cite{Yan}%
. A similar complex, albeit not $\mathcal{PT}$-symmetric, version of
Burgers' equations plays an important role in the study of two-dimensional
Yang-Mills theory with an SU(N) gauge group \cite{Neuberger1,Neuberger2}.
The models considered in \cite{Neuberger1,Neuberger2} become $\mathcal{PT}$%
-symmetric after a Wick rotation, i.e $t\rightarrow it$.

Let us now consider the $\mathcal{PT}$-symmetrically deformed Burgers'
equation 
\begin{equation}
u_{t}+uu_{x;\varepsilon }=i\kappa u_{xx;\mu }\text{\qquad \qquad with \ }%
\kappa ,\varepsilon ,\mu \in \mathbb{R},  \label{defBurg}
\end{equation}
where for the time being we allow two different deformation parameters $%
\varepsilon $ and $\mu $.

Our first objective is to test whether this set of equations passes the
Painlev\'{e} test. Following the method proposed in \cite{Pain1}, we
therefore assume that the solution of (\ref{defBurg}) acquires the general
form of the Painlev\'{e} expansion 
\begin{equation}
u(x,t)=\sum\limits_{k=0}^{\infty }\lambda _{k}(x,t)\phi (x,t)^{k+\alpha }.
\label{u}
\end{equation}
Here $\alpha \in \mathbb{Z}_{-}$ is the leading order singularity in the
limit $\phi (x,t)=(\varphi (x,t)-\varphi _{0})\rightarrow 0$, with $\varphi
(x,t)$ being an arbitrary analytic function characterizing the singular
manifold, $\varphi _{0}$ being an arbitrary complex constant which can be
utilized to move the singularity mimicking the initial condition and the $%
\lambda _{k}(x,t)$ are analytic functions, which have to be computed
recursively.

\subsubsection{Leading order terms}

As a starting point we need to determine all possible values for $\alpha $
by substituting the first term of the expansion (\ref{u}), that is $%
u(x,t)\rightarrow \lambda _{0}(x,t)\phi (x,t)^{\alpha }$, into (\ref{defBurg}%
) and reading off the leading orders. For the three terms in (\ref{defBurg})
they are $u_{t}\sim \phi ^{\alpha -1}$, $uu_{x;\varepsilon }\sim \phi
^{\alpha +\alpha \varepsilon -\varepsilon }$ and $u_{xx;\mu }\sim \phi
^{\alpha \mu -\mu -1}$. In order for a non-trivial solution to exist the
last two terms have to match each other in powers of $\phi $, which
immediately yields $\alpha =(\varepsilon -\mu -1)/(\varepsilon -\mu +1)$ $%
\in \mathbb{Z}_{-}$. Thus $\alpha =-1$ and $\varepsilon =\mu $ is the only
possible solution. This means we observe from the very onset of the
procedure that only the models in which all $x$-derivatives are deformed
with the same deformation parameter have a chance to pass the Painlev\'{e}
test. Therefore we can conclude already at this stage that one of the
deformations of (\ref{Burgers}) studied in \cite{Yan}, i.e. $\varepsilon =1$
and $\mu $ generic, can not pass the Painlev\'{e} test. Hence they do not
possess the Painlev\'{e} property and are therefore not integrable.

\subsubsection{Recurrence relations}

Substituting next the\ Painlev\'{e} expansion (\ref{u}) for $u(x,t)$ with $%
\alpha =-1$ into (\ref{defBurg}) with $\varepsilon =\mu $ gives rise to the
recursion relations for the $\lambda _{k}$ by identifying powers in $\phi
(x,t)$. We find 
\begin{equation}
\begin{array}{lr}
\text{at order }-(2\varepsilon +1)\text{: \ \ \ \ } & \lambda
_{0}+i2\varepsilon \kappa \phi _{x}=0, \\ 
\text{at order }-2\varepsilon \text{:} & \phi _{t}\delta _{\varepsilon
,1}+\lambda _{1}\phi _{x}-i\kappa \varepsilon \phi _{xx}=0, \\ 
\text{at order }-(2\varepsilon -1)\text{:} & \qquad \partial _{x}(\phi
_{t}\delta _{\varepsilon ,1}+\lambda _{1}\phi _{x}-i\kappa \varepsilon \phi
_{xx})=0,%
\end{array}
\label{order}
\end{equation}
such that 
\begin{equation}
\lambda _{0}=-i2\varepsilon \kappa \phi _{x},\qquad \lambda
_{1}=(i\varepsilon \kappa \phi _{xx}-\phi _{t}\delta _{\varepsilon ,1})/\phi
_{x}\qquad \text{and\qquad }\lambda _{2}\text{ is arbitrary.}  \label{rr}
\end{equation}
This means the number of free parameters, i.e. $\varphi _{0}$ and $\lambda
_{2}$, at our disposal equals the order of the PDE, such that (\ref{defBurg}%
) passes the Painlev\'{e} test provided the series (\ref{u}) makes sense and
we can determine all $\lambda _{j}$ with $j>2$. To compute the remaining $%
\lambda _{j}$ we need to isolate them on one side of the equation and those
involving $\lambda _{k}$ with $k<j$ on the other side. We expect to find
some recursion relations of the form 
\begin{equation}
g(j,\phi _{t},\phi _{x},\phi _{xx},\ldots )\lambda _{j}=f(\lambda
_{j-1},\lambda _{j-2},\ldots ,\lambda _{1},\lambda _{0},\phi _{t},\phi
_{x},\phi _{xx},\ldots ),  \label{rec}
\end{equation}
with $g$ and $f$ being some functions characteristic for the system under
consideration. We will not present here these recursion relations for
generic values of $\varepsilon $ as they are rather cumbersome and we shall
only present the first non-trivial deformation, that is the case $%
\varepsilon =2$.

\subsubsection{Resonances}

\label{reson}

For some particular values of $j$, say $j=r_{1},\ldots ,r_{\ell }$, we might
encounter that the function $g$ in (\ref{rec}) vanishes. Clearly this leads
to an inconsistency and a failure of the Painlev\'{e} test unless $f$ also
vanishes. In case this scenario occurs, it implies that the recursion
relation (\ref{rec}) does not fix $\lambda _{j}$ and the compatibility
conditions $g=f=0$ lead to $\ell $ so-called resonances $\lambda _{r_{i}}$
for $i=1,\ldots \ell $. When $\ell +1$ is equal to the order of the
differential equation we can in principle produce a general solution which
allows for all possible initial values. It might turn out that some missing
free parameters are located before the start of the expansion (\ref{u}),
i.e. at $j<0$, so-called negative resonances which can be treated following
arguments developed in \cite{Fordy}. When not enough additional free
parameters exist to match the order of the differential equation, the series
is still of Painlev\'{e} type and is called defective.

It is straightforward to determine all possible resonances by following a
standard argument. The first term in the expansion (\ref{u}) gives rise to
the leading order singularity which needs to be cancelled by some yet
unknown term in the expansion. Let us carry out the calculation for Burgers
equation. Using the expression for $\lambda _{0}$ from (\ref{rr}) and making
the ansatz 
\begin{equation}
\tilde{u}(x,t)=-2i\varepsilon \kappa \frac{\phi _{x}}{\phi }+\vartheta \phi
^{r-1},  \label{vv}
\end{equation}
we can compute all possible values of $r$ for which $\vartheta $ becomes a
free parameter. Substituting $\tilde{u}(x,t)$ into (\ref{defBurg}) and
reading off the terms of the highest order, i.e. $\phi ^{-2\varepsilon -1+r}$%
, we find the necessary condition 
\begin{equation}
i2^{\varepsilon -1}\varepsilon ^{\varepsilon }\vartheta (r+1)(r-2)\kappa
^{\varepsilon }\phi _{x}^{2\varepsilon }=0,
\end{equation}
for a resonance to exist. This yields precisely to two resonances, one at $%
r=2$, corresponding to the third equation in (\ref{order}), and the
so-called universal resonance at $r=-1$. This means also at higher order we
can not encounter any inconsistencies or possible breakdowns of the Painlev%
\'{e} test for any value of the deformation parameter $\varepsilon $.

\subsubsection{From the Painlev\'{e} test via Painlev\'{e} property to
integrability}

Once it is established that a PDE passes the Painlev\'{e} test one needs to
be cautious about the conclusions one can draw as it is only a necessary but
not sufficient condition for the Painlev\'{e} property. In case one can also
guarantee the convergence of the series the PDE possess the Painlev\'{e}
property, which is taken as very strong evidence for the equation to be
integrable. This step has only been carried out rigorously in very rare
cases, e.g. in \cite{Joshi1,Joshi2}. Here we establish the convergence for
one particular deformation.

\subsubsection{The $\protect\varepsilon =2$ deformation}

As already mentioned, the details of the recursion relation for generic
values of $\varepsilon $ are rather lengthy and we shall therefore only
present the case $\varepsilon =2$ explicitly. In that case the deformed
Burgers' equation (\ref{defBurg}) becomes 
\begin{equation}
u_{t}+iuu_{x}^{2}+2\kappa u_{x}u_{xx}=0  \label{e2}
\end{equation}
The substitution of the Painlev\'{e} expansion (\ref{u}) into (\ref{e2}) and
the subsequent matching of equal powers in $\phi $ then yields the recursion
relation

\begin{eqnarray}
&&i\lambda _{0}\phi _{x}^{2}\left\{ \lambda _{j}\left[ (2j-3)\lambda
_{0}-2i((j-5)j+4)\kappa \phi _{x}\right] +2\lambda _{0}\left( \lambda
_{0}+2i\kappa \phi _{x}\right) \delta _{0,j}\right\} =  \label{recu} \\
&+&\sum_{n,m=1}^{j}\left\{ \lambda _{j-m-n-2}\lambda _{m,x}\lambda
_{n;x}+(m-1)\lambda _{m}\phi _{x}\left[ (n-1)\lambda _{j-m-n}\lambda
_{n}\phi _{x}+2\lambda _{j-m-n-1}\lambda _{n;x}\right] \right\}  \notag \\
&+&\sum_{n=1}^{j-1}\left\{ 2\lambda _{0,x}\left[ (n-1)\lambda
_{j-n-1}\lambda _{n}\phi _{x}+\lambda _{j-n-2}\lambda _{n;x}\right]
-2\lambda _{0}\phi _{x}\left[ (n-1)\lambda _{j-n}\lambda _{n}\phi
_{x}+\lambda _{j-n-1}\lambda _{n;x}\right] \right.  \notag \\
&&-2i\kappa \left\{ \lambda _{j-n,x}\left[ \lambda _{n-3;xx}+(n-3)\left(
(n-2)\lambda _{n-1}\phi _{x}^{2}+2\lambda _{n-2,x}\phi _{x}+\lambda
_{n-2}\phi _{xx}\right) \right] \right.  \notag \\
&&+\left. \left. (j-n-1)\lambda _{j-n}\phi _{x}\left[ \lambda
_{n-2;xx}+(n-2)\left( (n-1)\lambda _{n}\phi _{x}^{2}+2\lambda _{n-1,x}\phi
_{x}+\lambda _{n-1}\phi _{\text{xx}}\right) \right] \right\} \right\}  \notag
\\
&+&2\lambda _{0,x}\left[ (j-5)j+6\right] \kappa \lambda _{j-1}\phi
_{x}^{2}+\lambda _{j-2}\left[ 2(j-3)\kappa \phi _{xx}+i\lambda _{0,x}\right]
\notag \\
&-&2\lambda _{0}\phi _{x}\left\{ \lambda _{j-1}\left[ (j-2)\kappa \phi
_{xx}+i\lambda _{0;x}\right] +\kappa \left[ \lambda _{j-2;xx}+2(j-2)\phi
_{x}\lambda _{j-1;x}\right] \right\}  \notag \\
&+&(j-4)\lambda _{j-3}\phi _{t}+\lambda _{j-4;t}+2\kappa \lambda _{0,x}\left[
\lambda _{j-3;xx}+2(j-3)\phi _{x}\lambda _{j-2;x}\right] ,  \notag
\end{eqnarray}
which is indeed of the general form (\ref{rec}). Having brought all $\lambda
_{j}$ with $j>k$ to the left hand side of (\ref{recu}), we may now
successively determine the $\lambda _{j}$ to any desired order. Starting
with the lowest value $j=0$ the equation (\ref{recu}) reduces to 
\begin{equation}
\lambda _{0}^{2}\phi _{x}^{2}\left( \lambda _{0}+i4\kappa \phi _{x}\right)
=0,
\end{equation}
which leads to $\lambda _{0}=-i4\kappa \phi _{x}$ and thus simply reproduces
the expression in (\ref{rr}) for $\varepsilon =2$. For $j=1$ the equation (%
\ref{recu}) simplifies to 
\begin{equation}
-\lambda _{0}^{2}\lambda _{1}\phi _{x}^{2}=2\lambda _{0}\phi _{x}\left[
i\kappa \lambda _{0}\phi _{xx}+\left( \lambda _{0}+i4\kappa \phi _{x}\right)
\lambda _{0;x}\right] ,
\end{equation}
such that $\lambda _{1}=i2\kappa \phi _{xx}/\phi _{x}$, which coincides with
(\ref{rr}) for $\varepsilon =2$. When $j=2$ the equation acquires the form 
\begin{eqnarray}
\lambda _{0}\lambda _{2}\phi _{x}^{2}\left( \lambda _{0}+4\sigma \phi
_{x}\right) &=&2\phi _{x}\lambda _{1;x}\lambda _{0}^{2}-\lambda
_{0;x}^{2}\lambda _{0}+2\lambda _{1}\phi _{x}\lambda _{0;x}-2i\kappa \phi
_{xx}\lambda _{0;x}-4i\kappa \phi _{x}\lambda _{0;x}^{2}  \notag \\
&&-2i\kappa \phi _{x}\left( \lambda _{0,xx}-2\phi _{x}\lambda _{1;x}\right)
\lambda _{0}.  \label{fff}
\end{eqnarray}
$\allowbreak $It is evident that the left hand side vanishes identically and
upon substitution of the values for $\lambda _{0}$ and $\lambda _{1}$. We
can verify that this also holds for the right hand side of (\ref{fff}), thus
leading to the first resonance at level 2 and therefore to an arbitrary
parameter $\lambda _{2}$. One may now continue in this fashion to compute
the expansion to any finite order, but before we embark on this task we make
a few further simplification.

As the singularity has to be a noncharacteristic analytic movable
singularity manifold, we employ the implicit function theorem and make a
further assumption about the specific form of $\lambda _{k}(x,t)=\lambda
_{k}(t)$ and $\phi (x,t)=x-\xi (t)$, with $\xi (t)$ being an arbitrary
function. Then the equation (\ref{recu}) simplifies to a much more
transparent form

\begin{eqnarray}
&&8\kappa ^{2}\left( 8\kappa \delta _{0,j}+i(j-2)(j+1)\lambda _{j}(t)\right)
=\sum_{n,m=1}^{j}i(1-m)(n-1)\lambda _{m}(t)\lambda _{j-m-n}(t)\lambda _{n}(t)
\label{ret} \\
&+&\sum_{n=1}^{j-1}\left[ 2\kappa (n-1)\left( n^{2}-n-j(n-2)+2\right)
\lambda _{j-n}(t)\lambda _{n}(t)\right] +(j-4)\lambda _{j-3}(t)\xi ^{\prime
}(t)-\lambda _{j-4}^{\prime }(t).~~~~  \notag
\end{eqnarray}
Solving this equation recursively leads to the Painlev\'{e} expansion 
\begin{equation}
u(x,t)=-\frac{4i\kappa }{\phi }+\lambda _{2}\phi +\frac{\xi ^{\prime }}{%
8\kappa }\phi ^{2}-\frac{i\lambda _{2}^{2}}{20\kappa }\phi ^{3}-\frac{%
i\lambda _{2}\xi ^{\prime }}{96\kappa ^{2}}\phi ^{4}+\mathcal{O}(\phi ^{5}).
\label{exp}
\end{equation}
Clearly we can use (\ref{ret}) to extend this expansion to any desired
order. For the ordinary Burgers equations, i.e. $\varepsilon =1$, there
exist a simple choice for the free parameters, which terminates the
expansion, such that one may generate B\"{a}cklund and Cole-Hopf
transformations in a very natural way. Unfortunately (\ref{exp}) does not
allow an obvious choice of this form. Taking for instance $\lambda _{2}=0$
yields the expansion

\begin{eqnarray}
u(x,t) &=&-\frac{4i\kappa }{\phi }+\frac{\xi ^{\prime }\phi ^{2}}{%
2^{3}\kappa }-\frac{i\xi ^{\prime }{}^{2}\phi ^{5}}{7\times 2^{8}\kappa ^{3}}%
+\frac{i\xi ^{\prime \prime }\phi ^{6}}{5\times 2^{9}\kappa ^{3}}-\frac{\xi
^{\prime }{}^{3}\phi ^{8}}{35\times 2^{13}\kappa ^{5}}-\frac{23\xi ^{\prime
}\xi ^{\prime \prime }\phi ^{9}}{385\times 2^{13}\kappa ^{5}}-\frac{\xi
^{(3)}\phi ^{10}}{135\times 2^{14}\kappa ^{5}}  \notag \\
&&+\frac{19i\xi ^{\prime 4}\phi ^{11}}{3185\times 2^{18}\kappa ^{7}}-\frac{%
51i\xi ^{\prime 2}\xi ^{\prime \prime }\phi ^{12}}{385\times 2^{19}\kappa
^{7}}-\frac{i\left( 43641\xi ^{\prime \prime 2}+16460\xi ^{\prime }\xi
^{(3)}\right) \phi ^{13}}{779625\times 2^{20}\kappa ^{7}}+\mathcal{O}(\phi
^{14}).  \label{zweii}
\end{eqnarray}%
Being even more specific and assuming a travelling wave solution, the
general form of the movable singularity is $\xi (t)=\omega t$, which gives 
\begin{eqnarray}
u(x,t) &=&-\frac{4i\kappa }{\phi }+\frac{\omega \phi ^{2}}{2^{3}\kappa }-%
\frac{i\omega ^{2}\phi ^{5}}{7\times 2^{8}\kappa ^{3}}-\frac{\omega ^{3}\phi
^{8}}{35\times 2^{13}\kappa ^{5}}+\frac{19i\omega ^{4}\phi ^{11}}{3185\times
2^{18}\kappa ^{7}}+\frac{\omega ^{5}\phi ^{14}}{3185\times 2^{21}\kappa ^{9}}
\notag \\
&&-\frac{561i\omega ^{6}\phi ^{17}}{2118025\times 2^{28}\kappa ^{11}}-\frac{%
93\omega ^{7}\phi ^{20}}{3328325\times 2^{32}\kappa ^{13}}+\frac{%
625011i\omega ^{8}\phi ^{23}}{53003575625\times 2^{38}\kappa ^{15}}  \notag
\\
&&+\frac{32971\omega ^{9}\phi ^{26}}{53003575625\times 2^{41}\kappa ^{17}}-%
\frac{1509727i\omega ^{10}\phi ^{29}}{11501775910625\times 2^{46}\kappa ^{19}%
}+\mathcal{O}(\phi ^{30}).
\end{eqnarray}%
Clearly we can carry on with this procedure to any desired order.

\paragraph{Convergence of the Painlev\'{e} expansion}

Having established that the deformed Burgers equations pass the Painlev\'{e}
test for any value of the deformation parameter $\varepsilon $, let us now
see whether the obtained series converges such we may conclude that these
equations also posses the Painlev\'{e} property. It suffices to demonstrate
this for some specific cases. Taking for this purpose $\lambda _{2}=0$, we
can express the expansion (\ref{zweii}) in the general form 
\begin{equation}
u(x,t)=-\frac{4i\kappa }{\phi }+\phi \sum_{n=1}^{\infty }\alpha _{n}\phi ^{n}
\label{ser}
\end{equation}
and employ Cauchy's root test, i.e. $\sum_{n=1}^{\infty }\gamma _{n}$
converges if and only if $\lim_{n\rightarrow \infty }\left\vert \gamma
_{n}\right\vert ^{1/n}\leq 1$, to establish the convergence of the series.
We can easily find an upper bound for the real and imaginary parts of $%
\alpha _{n}$%
\begin{equation}
\left\vert \func{Re}\alpha _{3n-\nu }\right\vert \leq \frac{\left\vert \func{%
Re}p_{3n-\nu }(\xi ^{\prime },\xi ^{\prime \prime },\xi ^{\prime \prime
\prime },\ldots )\right\vert }{2^{3n+4-\nu }\Gamma (\frac{3n-\nu }{2}%
)\left\vert \kappa \right\vert ^{2n-1}}\qquad \text{for }\nu =0,1,2,
\label{bound}
\end{equation}
where the $p_{n}(\xi ^{\prime },\xi ^{\prime \prime },\xi ^{\prime \prime
\prime },\ldots )$ are polynomials of finite order in $t$, that is $%
\sum\nolimits_{n=0}^{\ell }\omega ^{n}t^{n}$ with $\ell <\infty $ and $%
\omega \in \mathbb{C}$. The same expression holds when we the replace real
part by the imaginary part on both sides of the inequality. We should also
comment that this point of the proof is not entirely rigorous in the strict
mathematical sense as we have only verified the estimate (\ref{bound}) up to
order thirty. Approximating now the gamma function in (\ref{bound}) by
Stirling's formula as $n\rightarrow \infty $%
\begin{equation}
\Gamma \left( \frac{n}{2}\right) \sim \sqrt{2\pi }e^{-n/2}\left( \frac{n}{2}%
\right) ^{\frac{n-1}{2}}
\end{equation}
we obtain 
\begin{equation}
\lim_{n\rightarrow \infty }\left\vert \func{Re}\alpha _{3n-\nu }\right\vert
^{\frac{1}{2}}\sim \frac{\left\vert \func{Re}p_{3n-\nu }\right\vert ^{1/n}}{%
2^{3+\frac{4-\nu }{n}}(2\pi )^{\frac{1}{2n}}e^{-\frac{1}{2}}(\frac{3n-\nu }{2%
})^{\frac{1}{2}-\frac{1}{2n}}\left\vert \kappa \right\vert ^{2-\frac{1}{n}}}%
=0.
\end{equation}
The same argument holds for the imaginary part, such that the series (\ref%
{ser}) converges for any value of $\kappa $ and choices for $\xi (t)$
leading to finite polynomials $p_{n}(\xi ^{\prime },\xi ^{\prime \prime
},\xi ^{\prime \prime \prime },\ldots )$. It is straightforward to repeat
the same argument for $\lambda _{2}\neq 0$.

Alternatively we can identify the leading order term in (\ref{e2}) and
integrate the deformed Burgers equation twice. In this way we change the ODE
into an integral equation 
\begin{equation}
u(x,t)=2\kappa \left\{ g(t)+\int\nolimits_{x_{1}}^{x}d\hat{x}\left[ \frac{i}{%
2}+\frac{1}{u^{2}(\hat{x},t)}\left( f(t)+\int\nolimits_{x_{0}}^{\hat{x}}d%
\tilde{x}\frac{u_{t}(\tilde{x},t)}{u_{\tilde{x}}(\tilde{x},t)}\right) \right]
\right\} ^{-1},  \label{inte}
\end{equation}%
where $g(t),f(t)$ are some functions of integration. When discretising this
equation, i.e. taking the left hand side to be $u_{n+1}(x,t)$ and replacing
all the $u(x,t)$ on the right hand side of this equation by $u_{n}(x,t)$, we
may iterate (\ref{inte}) with $u_{0}(x,t)=-4i\kappa /[x-\xi (t)]$ as initial
condition and recover precisely the expansion (\ref{exp}). Exploiting the
Banach fixed point theorem one may also use (\ref{inte}) as a starting point
to establish the convergence of the iterative procedure and therefore the
Painlev\'{e} expansion, similarly as was carried out for instance in \cite%
{Joshi1,Joshi2}.

\paragraph{Reduction from PDE to ODE}

Making further assumptions on the dependence of $u(x,t)$ on $x$ and $t$ we
can reduce the PDE to an ODE, and attempt to solve the resulting equation by
integration. A common assumption is to require the solution to be of the
form of a travelling wave $u(x,t)=\zeta (z)=\zeta (x-vt)$ with $v$ being
constant. When $v$ is taken to be real, even solutions will be invariant
under the original $\mathcal{PT}$-symmetry. With this ansatz the deformed
Burgers' equation for $\varepsilon =2$ (\ref{e2}) acquires the form 
\begin{equation}
-v\zeta _{z}+i\zeta \zeta _{z}^{2}+2\kappa \zeta _{z}\zeta _{zz}=0.
\label{as}
\end{equation}%
When $\xi _{z}\neq 0$ we can re-write this equation as 
\begin{equation}
\frac{d}{dz}\left( c-vz+\frac{i}{2}\zeta ^{2}+2\kappa \zeta _{z}\right) =0,
\end{equation}%
which can be integrated to 
\begin{equation}
\zeta (z)=e^{i\pi 5/3}(2v\kappa )^{1/3}\frac{\tilde{c}Ai^{\prime }(\chi
)+Bi^{\prime }(\chi )}{\tilde{c}Ai(\chi )+Bi(\chi )}
\end{equation}%
with $c$, $\tilde{c}$ being constants, $\chi =e^{i\pi /6}(vz-c)(2v\kappa
)^{-2/3}$ and $Ai(\chi )$, $Bi(\chi )$ denoting Airy functions.

\subsection{Painlev\'{e} test for the $\mathcal{PT}$-symmetrically deformed
KdV-equation}

The KdV-equation was found to be $\mathcal{PT}$-symmetric and was the first
equation for which deformations have been studied \cite{BBCF,AFKdV}. Next we
investigate the $\mathcal{PT}$-symmetrically deformed version of the
KdV-equation with two different deformation parameters $\varepsilon $ and $%
\mu $ 
\begin{equation}
u_{t}-6uu_{x;\varepsilon }+u_{xxx;\mu }=0\text{\qquad \qquad with \ }%
\varepsilon ,\mu \in \mathbb{R}.  \label{defKdV}
\end{equation}
The case $\mu =1$ and $\varepsilon $ generic was considered in \cite{BBCF}
and the case\ $\varepsilon =1$ and $\mu $ generic was studied in \cite{AFKdV}%
.

\subsubsection{Leading order terms}

As in the previous section we substitute $u(x,t)\rightarrow \lambda
_{0}(x,t)\phi (x,t)^{\alpha }$ into (\ref{defKdV}) in order to determine the
leading order term. From $u_{t}\sim \phi ^{\alpha -1}$, $uu_{x;\varepsilon
}\sim \phi ^{\alpha +\alpha \varepsilon -\varepsilon }$ and $u_{xxx;\mu
}\sim \phi ^{\alpha \mu -\mu -2}$ we deduce $\alpha =(\varepsilon -\mu
-2)/(\varepsilon -\mu +1)$ $\in \mathbb{Z}_{-}$, such that the only solution
is $\alpha =-2$ with $\varepsilon =\mu $. This means neither the case $\mu
=1 $ and $\varepsilon $ generic nor the case\ $\varepsilon =1$ and $\mu $
generic can pass the Painlev\'{e} test, but the hitherto uninvestigated
deformation with $\varepsilon =\mu $ has at this point still a chance to
pass it.

\subsubsection{Recurrence relations}

Substituting the\ Painlev\'{e} expansion (\ref{u}) for $u(x,t)$ with $\alpha
=-2$ into (\ref{defKdV}) with $\varepsilon =\mu $ gives rise to the
recursion relations for the $\lambda _{k}$ by identifying powers in $\phi
(x,t)$. We compute 
\begin{equation}
\begin{array}{lr}
\text{order }-(3\varepsilon +2)\text{: \ \ } & \lambda _{0}=\frac{1}{2}%
\varepsilon (3\varepsilon +1)\phi _{x}^{2}, \\ 
\text{order }-(3\varepsilon +1)\text{:} & \lambda _{1}=-\frac{1}{2}%
\varepsilon (3\varepsilon +1)\phi _{xx}, \\ 
\text{order }-3\varepsilon \text{:} & \qquad \lambda _{2}=\frac{\varepsilon
(3\varepsilon +1)}{24}\left( \frac{4\phi _{x}\phi _{xxx}-3\phi _{xx}^{2}}{%
\phi _{x}^{2}}\right) +\delta _{\varepsilon ,1}\frac{\phi _{t}}{6\phi _{x}},
\\ 
\text{order }-(3\varepsilon -1)\text{:} & \lambda _{3}=\frac{\varepsilon
(3\varepsilon +1)}{24}\left( \frac{4\phi _{x}\phi _{xx}\phi _{xxx}-3\phi
_{xx}^{3}-\phi _{x}^{2}\phi _{4x}}{\phi _{x}^{4}}\right) +\delta
_{\varepsilon ,1}\frac{\phi _{t}\phi _{xx}-\phi _{x}\phi _{xt}}{6\phi
_{x}^{3}}, \\ 
\text{order }-(3\varepsilon -2)\text{:} & \lambda _{4}=\frac{\varepsilon
(3\varepsilon +1)}{24}\left( \frac{6\phi _{x}\phi _{xx}^{2}\phi _{xxx}-\frac{%
15}{4}\phi _{xx}^{4}-\frac{3}{2}\phi _{x}^{2}\phi _{xx}\phi _{4x}}{\phi
_{x}^{6}}+\frac{\phi _{x}\phi _{5x}-5\phi _{xxx}^{2}}{5\phi _{x}^{4}}\right)
.%
\end{array}
\label{recur}
\end{equation}%
We find that the relation at order $-(3\varepsilon -2)$ becomes an identity
only for $\varepsilon =1$, which makes us suspect that also at higher order
we will not encounter compatibility conditions and therefore will not have
enough parameters equaling the order of the differential equation. To test
whether new compatibility conditions arise at higher levels we can use the
same general argument as in subsection \ref{reson}.

\subsubsection{Resonances}

We try once again to match the first term in the expansion (\ref{u}) with
some term of unknown power. Using the expression for $\lambda _{0}$ in (\ref%
{recur}) and making the ansatz 
\begin{equation}
\tilde{u}(x,t)=\frac{1}{2}\varepsilon (3\varepsilon +1)\frac{\phi _{x}^{2}}{%
\phi ^{2}}+\vartheta \phi ^{r-2},
\end{equation}
we compute all possible values of $r$ for which $\vartheta $ becomes a free
parameter. Substituting $\tilde{u}(x,t)$ into (\ref{defKdV}) and reading off
the terms of the highest order, i.e. $\phi ^{-3\varepsilon -2+r}$, we find
the necessary condition 
\begin{equation}
\varepsilon ^{\varepsilon }(-i)^{\varepsilon -1}(3\varepsilon
+1)^{\varepsilon -1}(r+1)\left[ 6(1+3\varepsilon )-2(2+3\varepsilon )r+r^{2}%
\right] \vartheta \phi _{x}^{3\varepsilon }=0,
\end{equation}
for a resonance to exist. We observe the presence of the universal resonance
at $r=-1$. The bracket containing the quadratic term in $r$ can be
factorized as $(r-r_{-})(r-r_{+})$ with $r_{\pm }=-(2+3\varepsilon )\pm 
\sqrt{9\varepsilon ^{2}-6\varepsilon -2}$, such that $r_{\pm }\in \mathbb{Z}$
for $9\varepsilon ^{2}-6\varepsilon -2=n^{2}$ with $n\in \mathbb{N}$. For
the solution of this equation $\varepsilon _{\pm }=(1\pm \sqrt{n^{2}+3})/3$
to be an integer we need to solve a Diophantine equation $3+n^{2}=m^{2}$
with $n,m\in \mathbb{N}$, which only admits $n=1$ and $m=2$ as solution.
Thus the bracket only factorises in the case $\varepsilon =1$ into $%
(r-6)(r-4)$. Hence, only in that case the system can fully pass the Painlev%
\'{e} test. Nonetheless, we may still be able to obtain a defective series
if all remaining coefficients $\lambda _{j}$ may be computed recursively.
This is indeed the case as we demonstrate in detail for one particular
choice of the deformation parameter.

\subsubsection{$\protect\varepsilon =2$ deformation}

For $\varepsilon =\mu =2$ the deformed KdV equation (\ref{defKdV}) acquires
the form 
\begin{equation}
u_{t}-6iuu_{x}^{2}+2iu_{xx}^{2}+2iu_{x}u_{xxx}=0
\end{equation}%
Since the expression become rather lengthy for generic values in the
expansion we will present here only the case $\lambda _{k}(x,t)=\lambda
_{k}(t)$ and $\phi (x,t)=x-\xi (t)$, with $\xi (t)$ being an arbitrary
function. We find a recursion relation of the form (\ref{rec}) 
\begin{eqnarray}
&&-28i(1+j)(j^{2}-16j+42)\lambda
_{j}(t)=-6i\sum_{n=1}^{j}\sum_{m=1}^{j-n-1}\left\{ (m-2)(n-2)\lambda
_{m}(t)\lambda _{n}(t)\lambda _{j-m-n}(t)\right\}   \notag \\
&&+2i\sum_{n=1}^{j-1}\left\{
[(7-k)n^{3}+(k-4)kn^{2}+(18-5k)kn+6k(5+k)-28(6+n)]\lambda _{j-n}(t)\lambda
_{n}(t)\right\}   \notag \\
&&+\lambda _{j-6}^{\prime }(t)+(j-7)\lambda _{j-5}^{\prime }(t).
\end{eqnarray}%
The recursive solution of this equation leads to the expansion 
\begin{eqnarray}
u(x,t) &=&\frac{7}{\phi ^{2}}+\frac{i\xi ^{\prime }\phi ^{3}}{156}+\frac{%
(\xi ^{\prime })^{2}\phi ^{8}}{192192}-\frac{\xi ^{\prime \prime }\phi ^{9}}{%
681408}+\frac{i(\xi ^{\prime })^{3}\phi ^{13}}{73081008}-\frac{725i\xi
^{\prime }\xi ^{\prime \prime }\phi ^{14}}{216449705472}+\frac{i\xi ^{\prime
\prime \prime }\phi ^{15}}{20262348288}  \notag \\
&&-\frac{340915(\xi ^{\prime })^{4}\phi ^{18}}{23989859332927488}+\frac{%
1867(\xi ^{\prime })^{2}\xi ^{\prime \prime }\phi ^{19}}{758331543121152}+%
\mathcal{O}(\phi ^{20}).
\end{eqnarray}%
Thus we have obtained a solution of Painlev\'{e} type for the deformed KdV
equation, albeit without enough free parameters, i.e. without the
possibility to accommodate all possible initial values. This means we have a
so-called defective series. As in the case of the deformed Burgers equation
it is instructive to consider the series for travelling wave solutions, i.e.
taking $\xi (t)=\omega t$, which yields 
\begin{eqnarray}
u(x,t) &=&\frac{7}{\phi ^{2}}+\frac{i\omega \phi ^{3}}{156}+\frac{\omega
^{2}\phi ^{8}}{192192}+\frac{i\omega ^{3}\phi ^{13}}{73081008}-\frac{%
340915\omega ^{4}\phi ^{18}}{23989859332927488}+\frac{391907i\omega ^{5}\phi
^{23}}{56760007181706436608}  \notag \\
&&-\frac{38892808841\omega ^{6}\phi ^{28}}{507260097462393341102260224}+%
\mathcal{O}(\phi ^{33}).
\end{eqnarray}%
Clearly we can carry on with this analysis to any desired order. The
convergence of the expansion can be established in a similar fashion as we
demonstrated for Burgers equation in the previous subsection or by making
use of an integral equation of the type (\ref{inte}). We find a similar
behaviour for other values of $\varepsilon $. 

\section{Conclusion}

We have carried out the Painlev\'{e} test for $\mathcal{PT}$-symmetric
deformations of the Burgers equation and the KdV equation. When deforming
both terms involving space derivatives, we found that the deformations of
the Burgers equation pass the test. In specific cases we have also
established the convergence of the series, such that these equations have in
addition the Painlev\'{e} property. Based on the conjecture by Ablowitz,
Ramani and Segur we take this as very strong evidence that these equations
are integrable. Regarding these models as new integrable systems leads
immediately to a sequence of interesting new problems related to features of
integrability, which we intend to address in a future publication \cite{PAF}%
. It is very likely that these systems admit soliton solutions and it should
be possible to compute the higher charges by means of Lax pairs, Dunkl
operators or other methods. We should point out that most of our arguments
will still hold when we start in (\ref{defBurg}) with the usual Burgers
equation, which has broken $\mathcal{PT}$-symmetry, i.e. with $\sigma
=i\kappa \in \mathbb{R}$. However, when embarking on the computation of
charges and in particular energies we expect to find a severe difference as
then the $\mathcal{PT}$-symmetry has a bearing on the reality of the
eigenvalues of the charges.

For the KdV equation our findings suggest that their $\mathcal{PT}$%
-symmetric deformations are not integrable, albeit they allow for the
construction of a defective series.

In future work one could also include deformations of the term involving the
time derivative and it would clearly be very interesting to investigate
other $\mathcal{PT}$-symmetrically integrable systems in the manner in order
to establish their integrability.

\medskip \noindent \textbf{Acknowledgments}: P.E.G.A. is supported by a City
University London research studentship.


\end{document}